\documentclass[aps,twocolumn,tightenlines,amsmath,amssymb,superscriptaddress]{revtex4-2}
\usepackage{graphicx}
\usepackage{color}
\usepackage[colorlinks=true]{hyperref}
\hypersetup{citecolor=black, urlcolor=black, linkcolor=black}
\usepackage{ulem}
\graphicspath{ {./images/} }

\begin{document}
\title{Non-classical excitation of a solid-state quantum emitter}
\author{L.~M.~Hansen} 
\email{lena.maria.hansen@univie.ac.at}
\affiliation{University of Vienna, Faculty of Physics, Vienna Center for Quantum Science and Technology (VCQ), 1090 Vienna, Austria}
\affiliation{Christian Doppler Laboratory for Photonic Quantum Computer, Faculty of Physics, University of Vienna, Vienna, Austria}
\author{F.~Giorgino}
\affiliation{University of Vienna, Faculty of Physics, Vienna Center for Quantum Science and Technology (VCQ), 1090 Vienna, Austria}
\affiliation{Christian Doppler Laboratory for Photonic Quantum Computer, Faculty of Physics, University of Vienna, Vienna, Austria}
\author{L.~Jehle}
\affiliation{University of Vienna, Faculty of Physics, Vienna Center for Quantum Science and Technology (VCQ), 1090 Vienna, Austria}
\author{L.~Carosini}
\affiliation{University of Vienna, Faculty of Physics, Vienna Center for Quantum Science and Technology (VCQ), 1090 Vienna, Austria}
\affiliation{Christian Doppler Laboratory for Photonic Quantum Computer, Faculty of Physics, University of Vienna, Vienna, Austria}
\author{J.~C.~L\'{o}pez~Carre\~{n}o}
\affiliation{Institute of Theoretical Physics, University of Warsaw, ul. Pasteura 5, 02-093 Warsaw, Poland}
\author{I.~Arrazola}
\affiliation{Instituto de F\'isica Te\'orica, UAM-CSIC, Universidad Aut\'onoma de Madrid, Cantoblanco, 28049 Madrid, Spain}
\affiliation{Vienna Center for Quantum Science and Technology, Atominstitut, TU Wien, 1040 Vienna, Austria}
\author{P.~Walther}
\email{philip.walther@univie.ac.at}
\affiliation{University of Vienna, Faculty of Physics, Vienna Center for Quantum Science and Technology (VCQ), 1090 Vienna, Austria}
\affiliation{Christian Doppler Laboratory for Photonic Quantum Computer, Faculty of Physics, University of Vienna, Vienna, Austria}
\affiliation{ University of Vienna, Research Network for Quantum Aspects of Space Time (TURIS), 1090 Vienna, Austria}
\affiliation{Institute for Quantum Optics and Quantum Information (IQOQI) Vienna, Austrian Academy of Sciences, Vienna, Austria}
\author{J.~C.~Loredo}
\email{juan.loredo1@gmail.com}
\affiliation{University of Vienna, Faculty of Physics, Vienna Center for Quantum Science and Technology (VCQ), 1090 Vienna, Austria}
\affiliation{Christian Doppler Laboratory for Photonic Quantum Computer, Faculty of Physics, University of Vienna, Vienna, Austria}

\begin{abstract}{The interaction between a single emitter and a single photon is a fundamental aspect of quantum optics. This interaction allows for the study of various quantum processes, such as emitter-mediated single-photon scattering and effective photon-photon interactions. However, empirical observations of this scenario and its dynamics are rare, and in most cases, only partial approximations to the fully quantized case have been possible. Here, we demonstrate the resonant excitation of a solid-state quantum emitter using quantized input light. For this light-matter interaction, with both entities quantized, we observe single-photon interference introduced by the emitter in a coherent scattering process, photon-number-depended optical non-linearities, and stimulated emission processes involving only two photons. We theoretically reproduce our observations using a cascaded master equation model. Our findings demonstrate that a single photon is sufficient to change the state of a solid-state quantum emitter, and efficient emitter-mediated photon-photon interactions are feasible. These results suggest future possibilities ranging from enabling quantum information transfer in a quantum network to building deterministic entangling gates for photonic quantum computing.}
\end{abstract}
\maketitle

Studying light interacting with matter at the most fundamental level offers a number of possibilities for the development of quantum technologies~\cite{cQED:Rempe15,qNet:Lodahl19}. On the one hand, non-classical states of light are suitable as carriers and for processing quantum information~\cite{photRev:Sciarrino18,photRev:Pryde19} because they virtually do not interact with their environment. However, the lack of direct photon-photon interactions also prevents access to, e.g., deterministic entangling gates on quantized states of light---an obstacle for building larger scalable quantum photonic systems.

Light strongly interacts with resonant atomic systems. For instance, coherent scattering involves the absorption of one photon subsequent re-emission of a single photon; it has been theoretically shown that even one photon can saturate an atomic transition~\cite{1pump:Leuchs09,1pump:Fan10,1pump:Scarani11,Npump:Combes12}. Moreover, an atomic system is inherently quantum non-linear, mediating various phenomena of effective photon-photon interactions, such as electromagnetically induced transparency~\cite{EIT:Marangos05,EIT:Roy11}, photon-blockade~\cite{blockade:Kimble05,blockade:Bachelard20,ZubizarretaCasalengua.2020}, and entangling quantum logic operations~\cite{Kim:2013aa,Hacker:2016aa}.

	\begin{figure*}[htp!]
		\centering
		\includegraphics[width=1\textwidth]{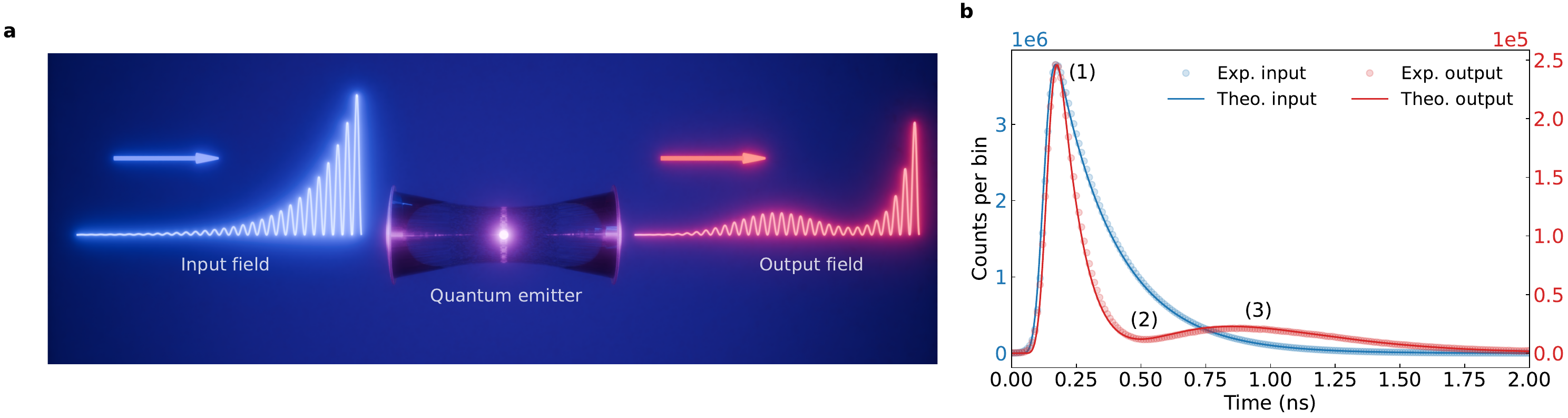}\vspace{0mm}	
		\caption{\textbf{Single-photon interference mediated by a quantum emitter.} \textbf{a}, A single-photon wave packet is sent as input to a QD coupled to a cavity. The light-matter interaction between the quantized light field and the quantum emitter is revealed in the temporal profile of the output wave packet that features single-photon interference. The process involves the quantum interference of two scenarios: attenuated propagation of the input field and re-emission after absorption. \textbf{b}, Experimentally, we investigate this concept by generating a single photon from the emitter and sending it back again to the emitter as quantized input field (blue dots). As predicted, the observed output field (red dots) is governed by (1) the transmitted part of the photon wave packet, (2) destructive interference, and (3) a second maxima indicating the re-emission of a photon. Due to the high single-photon count rate the Poissonian error bars are not visible in the plot. The observations are confirmed by the results of theoretical simulations (solid lines) that follow the cascaded master equation approach.}
	\label{fig:1}
	\end{figure*}
 
However, empirically observing an atomic system interacting with a single photon has remained a formidable challenge, as one requires a sufficiently controlled and strongly coupled quantum light-matter platform. Although scarce, a few experiments have reported aspects of quantized input light interacting with an atomic system, for instance, a quantum network of neutral atoms interconnected via single photons~\cite{Ritter:2012aa}, single photons scattered by a single Rubidium atom~\cite{Leong:2016aa}, heralded single-photon absorption in a trapped ion system~\cite{Piro:2011aa}, and electronic energy transfer in a photosynthetic complex induced by a single photon~\cite{Li:2023aa}. Despite impressive demonstrations, these previous works were limited by weak signals, which strongly hindered their ability to investigate the rich photon dynamics and statistics inherent in the fundamental process.

A solid-state platform offers a promising alternative for studying coherent quantum light-matter interactions. Significant progress has been made in both cavity and waveguide quantum electrodynamics~\cite{cavityQED:Walther06,wgQED:Poddubny23}. In particular, semiconductor quantum dots (QDs) coupled to resonant cavities and photonic crystal waveguides~\cite{sps:senellart17,sps:Lodahl21} constitute a leading candidate for the generation of various types of quantum light: highly-efficient sources of single and entangled photons~\cite{SPS:Somaschi16,SPS:Xing16,entpairs:liu19,Tomm:2021aa}, quantum superpositions and entangled states of photon-numbers~\cite{pnumbers:loredo19,pnent:wein22}, and deterministically generated squeezed states~\cite{intSqz:Pan20,quadSqz:Magro23}. Moreover, these solid-state systems led to multiple demonstrations of optical non-linearities involving a small number of photons on average, with works including photon-number dependent spatial~\cite{Bennett:2016aa,DeSantis:2017aa,Javadi:2015aa,sorting:Lodahl21,Hallett:18} and temporal~\cite{Tomm:2023aa,phasenonlinear:Jeannic23} ordering. However, these previous works employed classical weak coherent input light, thus leaving non-classical light-matter interactions in a solid-state platform without observation.

Here, we demonstrate a quantum light-matter interface in the solid-state. A single QD coupled to an optical cavity is resonantly excited with non-classical light. When the input light consists of a single photon, the output field stems from the quantum interference between the re-emitted part of the wave packet and the incoming part of the wave packet of the single photon. We demonstrate that this process is coherent by performing first- and second-order correlation measurements; indeed, the time-evolving output signal exhibits a point of destructive interference, and consecutive output wave packets display Hong-Ou-Mandel (HOM) two-photon interference~\cite{HOM87}. Moreover, we observe optical photon-number depended non-linearities occurring for quantized input light, as revealed by changes in the photon-number statistics. Further, we find signatures of stimulated emission processes between two photons, verified by analyzing the temporal mode of the generated two-photon Fock state.
\subsection*{Atom-mediated single-photon interference} 
Figure~\ref{fig:1}a depicts the general concept of this work. Here, a quantum emitter, an InAs/GaAs QD in a cavity, is excited with a non-classical wave packet. The output signal resulting from the quantum light-matter interaction is then collected for further analysis.

	\begin{figure*}[htp!]
		\centering
		\includegraphics[width=0.98\textwidth]{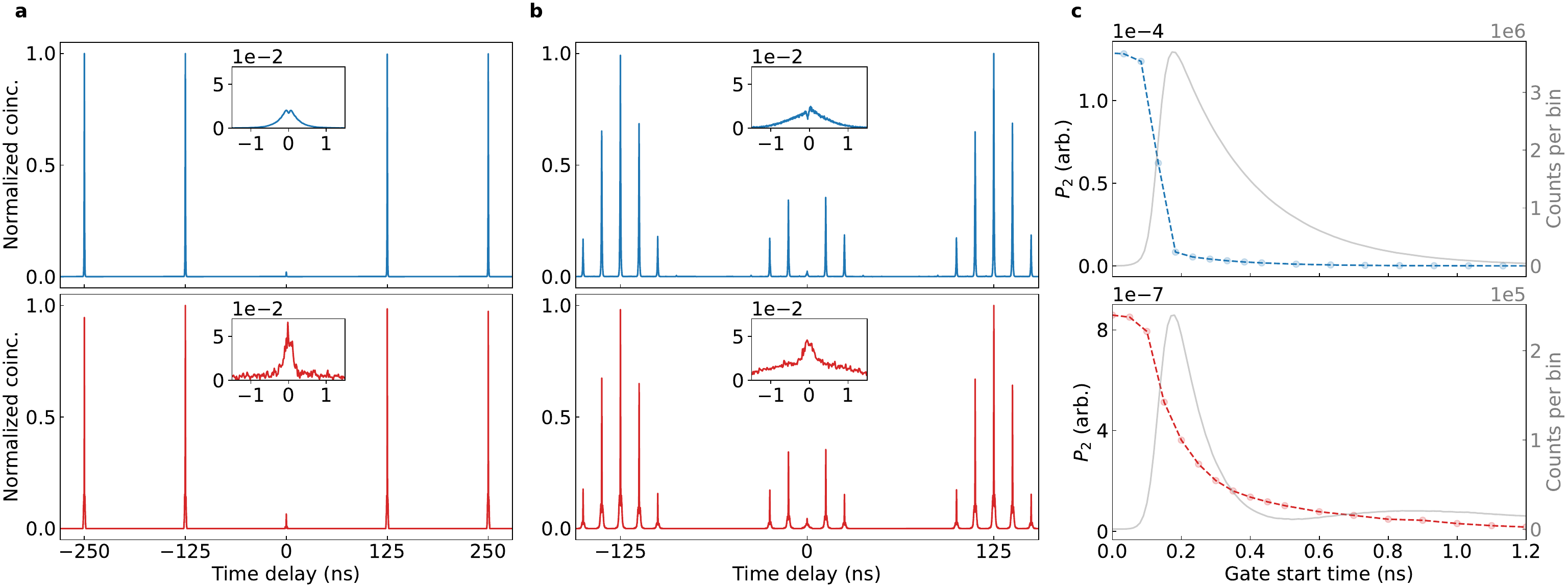}\vspace{-2mm}	
		\caption{\textbf{Measurements of single-photon purity and indistinguishability.} \textbf{a}, Comparing the photon statistics of the input field (top) and the output field (bottom) reveals that single-photon components are suppressed in the output field relative to multiphoton components. \textbf{b}, Our observations of the indistinguishability of the quantized input fields (top) compared to the scattered fields (bottom) indicate that indistinguishability remains in the output field, validating the coherence of the process. \textbf{c}, Time-gated two-photon probabilities $P_2$ estimated at the detector level for different starting gate times indicate a different temporal structure of two-photon components along the lifetime in the output field. The temporal histogram of each signal is shown in gray as a reference.} 
	\label{fig:2}
	\end{figure*}
	
In our experiment, the input non-classical light consists of resonance fluorescence from the same QD. That is, we first prepare a highly efficient single-photon source~\cite{h3GHZ:Cao23} (see Methods) and then use a setup with active path control that allows us to choose between two options: either the signal is back-reflected to the emitter or it is sent to the detection stage. With this setup, we first send the resonance fluorescence signal back toward the QD and later collect the resulting output signal; see the Supplementary Information for details on the experimental setup. 

We can tune the input light statistics by choosing the pulse area of the excitation laser that generates the resonance fluorescence signal, from mostly single photons---at $\pi$-pulse excitation---to states with one- and two-photon terms---e.g., at $2\pi$-pulse excitation~\cite{pnumbers:loredo19,Fischer:2017zr}. We first study the resulting time dynamics with an input signal generated by $\pi$-pulses. In this scenario, the quantum emitter interacts with a resonant single-photon wave packet. As the wave packet reaches the emitter, part of it is transmitted, and the excitation probability of the emitter begins to increase. During the interaction the emission probability keeps increasing, leading to quantum interference between the incoming wave packet and the reemitted part. Once the input wave packet duration ends, the emitter retains the probability of being excited and spontaneously decays without interference. Hence, the observation of destructive interference in the temporal output profile is due to coherent single-photon scattering~\cite{Npump:Combes12,coh1p:Chen11}. 

The optimal input temporal shape for achieving atomic inversion is formed as an exponentially increasing curve~\cite{1pump:Leuchs09,1pump:Fan10}---the time-reversed profile of the spontaneous emission decay. In our case, however, we see the opposite scenario because we back-reflect resonance fluorescence: a single photon arrives at the artificial atom with a field amplitude that first rises rapidly and then exponentially decays. Figure~\ref{fig:1}b shows the measured input time profile with a lifetime of $\tau\,{=}\,227$~ps, and the time profile of the output field after the interaction with the emitter. We observe the predicted atom-mediated single-photon interference phenomenon and identify three main regions: (1) an earlier part governed by an undisturbed signal---i.e., the input photon is not absorbed, it enters the optical cavity, and it is released back again into the collection mode; (2) an intermediate part displaying a point of minimal intensity given by destructive interference; and (3) a final part governed by spontaneous emission of an atom subject to an exponentially-decaying input field. Our measurements are confirmed by theoretical simulations that follow a cascaded master equation approach in which two identical subsystems are considered, with the output of the first system serving as input for the second system (see Methods).
\subsection*{Input-output photon statistics}
We now describe the photon statistics of the input and output fields for the scenario described above. We first measure the single-photon purity of the input field by using our active path-controlled setup and choosing to directly collect the resonance fluorescence at $\pi$-pulse excitation without sending it back to the artificial atom. As shown in Fig.~\ref{fig:2}, a time-gated Hanbury Brown and Twiss second-order autocorrelation measurement yields the value $g^{(2)}_{\text{in}}(0)\,{=}\,2.540(2)\%$, and an HOM experiment on consecutive emissions yields an indistinguishability value~\cite{HOM:Ollivier21} of $\mathcal{I}_\text{in}\,{=}\,91.90(5)\%$. 
These measurements demonstrate the high fidelity of the single photon used as an input in the quantum light-matter interface. We note that the multi-photon component in this state is caused by re-excitation of the QD driven by ${\sim}25$~ps long pulses---shorter pump pulses result in higher purities (see Supplementary Information). However, we deliberately choose this pulse length to study scenarios with stronger varying photon statistics along Rabi oscillations, which we describe below. We now perform the same measurements on the output field resulting from the atom-light interaction and find a higher value of $g^{(2)}_\text{out}(0)\,{=}\,6.32(4)\%$, as well as a decrease in the output field indistinguishability $\mathcal{I}_\text{out}\,{=}\,75.0(4)\%$.

	\begin{figure*}[htp!]
		\centering
		\includegraphics[width=1\textwidth]{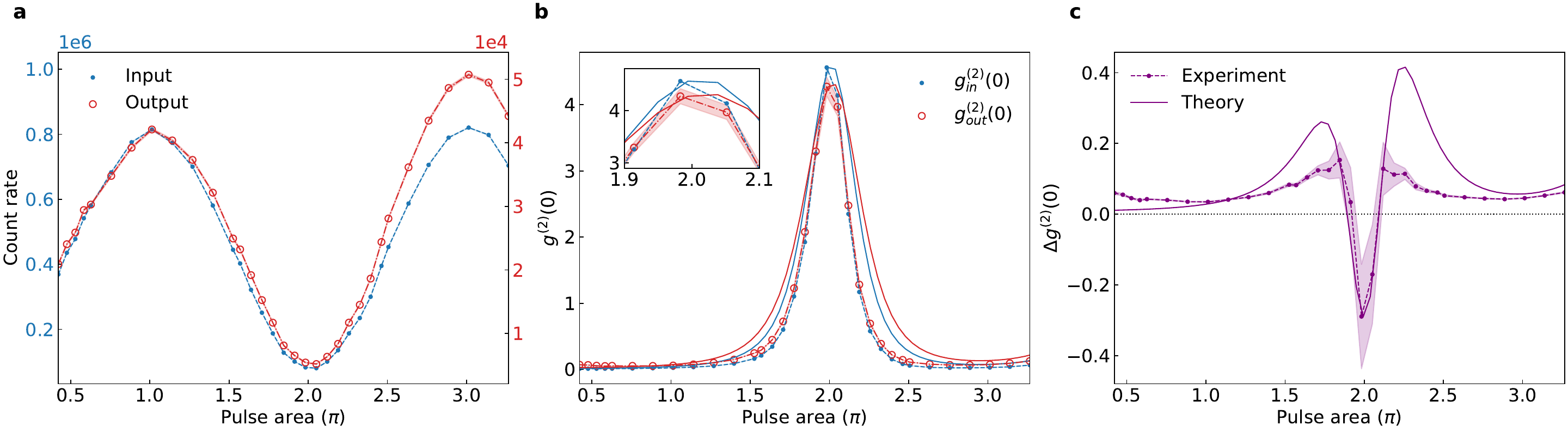}\vspace{-2mm}	
		\caption{\textbf{Photon statistics.} \textbf{a}, Characteristic Rabi oscillations of the non-classical input field for resonant laser excitation are shown in blue, and the output intensity resulting from non-classical excitation is plotted in red. \textbf{b}, The zero-delay values of the second-order autocorrelation for the input (blue dots) are measured for increasing pulse areas of the driving laser field, such that the photon statistic of the input field is tuned. The non-linear interaction between the input field and the QD causes a change in the measured output photon statistics (red dots). The solid lines are theoretical simulations of these measurements. Error bars are estimated following Poissonian counting statistics and are not visible for the results of the input field. \textbf{c}, The output field contains overall increased multiphoton components compared to the input field, and the difference $\Delta g^{(2)}(0)\,{=}\,g^{(2)}_\text{out}(0){-}g^{(2)}_\text{in}(0)$ is positive except for points around $2\pi$-pulse excitation that generates the input.} 
	\label{fig:3}
	\end{figure*}

The lower single-photon purity in the output field results from single-photon interference and effective photon-photon interactions, leading to a higher ratio of collected photon flux for multi-photon states~\cite{Npump:Combes12, sorting:Lodahl21}. Our results experimentally verify these effects for a non-classical input state. Our findings presented in Fig.~\ref{fig:2}b are evidence of the coherence of this process, showing for the first time that the output field maintains indistinguishability, an important conclusion for network applications.

The origin of multi-photon components is fundamentally different for the input and the output field. For the input field, these components are created from re-excitation. That is, with a small probability, part of the driving classical pulse area produces a first photon, leaving most of the pulse area available to re-populate the excited state. Therefore, a second photon is produced with the remaining pulse area, with this photon contributing primarily to the high input single-photon purity. Note that in this scenario the two photons occupy orthogonal time-modes along the input temporal profile. However, the multi-photon components of the output field can exhibit a different temporal structure depending on the interaction with the emitter.

This can be tested by time-gated photon statistics measurements and calculating the resulting upper bound for two-photon contribution. For simplicity, we consider multi-photon components up to two-photon terms since they are dominant compared to higher multi-photon contributions. Figure~\ref{fig:2}c (top) displays the upper-bound of the two-photon component of the input field for increasing gate start times. The main feature of this analysis is the rapid decrease in the amount of the two-photon component soon after gating begins. That is, by neglecting the photon statistics within the signal rise-time, the two-photon component of the remaining field is largely suppressed; thus, the remaining state is much closer to one (and only one) photon. The same gated measurement on the output field leads to a different conclusion, as shown in Fig.~\ref{fig:2}c (bottom). Here, we do not observe a sharp decrease in the relative two-photon component. In contrast, the amount of two-photon terms, decreases more slowly, which is consistent with the decrease in intensity of the entire wave packet. Thus, we observe two-photon components in the output field at times when they were absent in the input field. In the input field, a second photon is created only during the duration of the laser field due to re-excitation. However, in the output field, two photons can be observed due to temporal restructuring caused by the scattering dynamics from the emitter, which we will investigate further in the following section.
\subsection*{Effective photon-photon interaction} 
We now perform a scan of the classical pulse area responsible for generating the input non-classical light. The resulting Rabi oscillations are shown in Fig.~\ref{fig:3}a, together with the intensity of the output field stemming from the respective atom-light interaction. We note an increase in the output intensity for higher multi-photon contributions along the Rabi oscillations. This is consistent with theoretical predictions~\cite{Npump:Combes12,2ppump:Fan07,2ppump:Baranger10}, where higher photon numbers lead to higher photon flux in the output field.

	\begin{figure*}[htp!]
		\centering
		\includegraphics[width=1\textwidth]{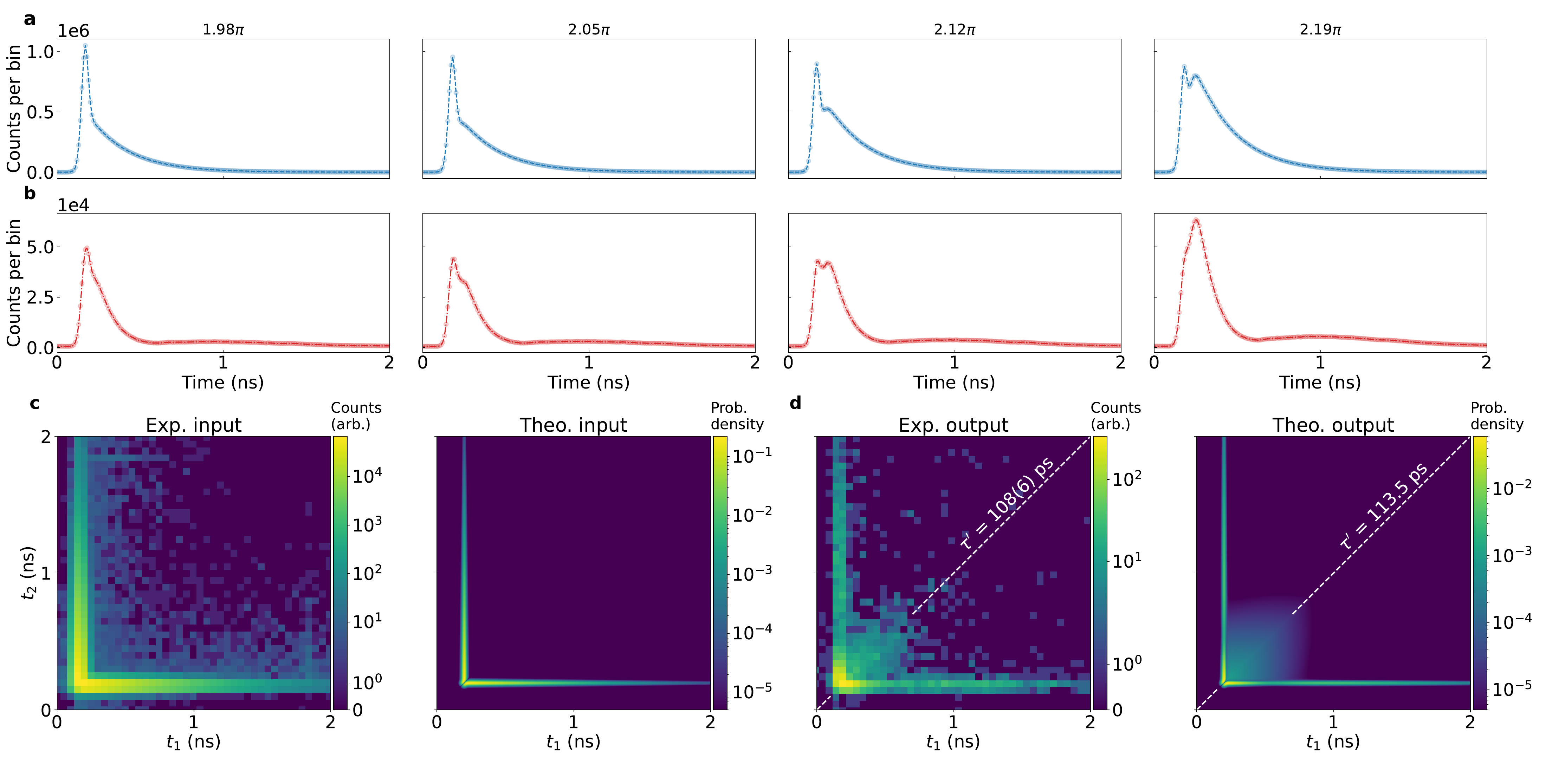}\vspace{-5mm}	
		\caption{\textbf{Evidence of effective photon-photon interactions.} \textbf{a}, Temporal histograms of the input signal for driving pulse areas around $2\pi$-pulse laser excitation showing signatures of two-photon emission in separate temporal modes. \textbf{b}, The resulting output signal has an increase in counts for later time bins. \textbf{c}, Measured and theoretical time-resolved intensity correlation maps of the input signal for $2\pi$-pulse laser excitation evidence that two photons within the input field do not occur simultaneously. The temporal axes of the measured time-resolved intensity correlation maps have an experimental resolution of $50~\text{ps}$. \textbf{d}, The corresponding measured and theoretical maps of the output signal, revealing simultaneous correlations along the diagonal ($t_1\,{=}\,t_2$) as a signature of stimulated emission. This is an observation of generating correlated two-photon states from two separate single-photon states.}
	\label{fig:4}
	\end{figure*}

Moreover, second-order autocorrelation measurements along the Rabi oscillations, as shown in Fig.~\ref{fig:3}b, exhibit an intriguing behavior: the output $g_{\text{out}}^{(2)}(0)$ values are higher than the input $g_{\text{in}}^{(2)}(0)$ values, except around $2{\pi}$-pulse excitation, where the two-photon component of the input non-classical field is higher~\cite{pnumbers:loredo19,Fischer:2017zr}. Indeed, Fig.~\ref{fig:3}c displays the difference between these measurements, clearly showing the change in the sign near the excitation using $2{\pi}$-pulses. This is a consistent behavior that we observe also for other values of the excitation pulse length that generate the input non-classical field (see Supplementary Information). The observed feature is a result of the non-linear interaction between the emitter and the input field. The incoming temporally separated two-photon components are partially changed to one-photon components in the detected output field, leading to lower $g_{\text{out}}^{(2)}(0)$ values. We note that linear transformations, such as unitary evolutions or optical losses, are not able to modify these photon statistics within the light field. Therefore, these measurements demonstrate the non-linear nature of the light-matter interface presented in this work.

A particularly interesting scenario occurs with the input state generated near a $2\pi$-pulse excitation, where we can study instances involving two photons arriving at the artificial atom. Now, the input field consists of one photon that appears first and that is temporally shorter and a second photon with a longer time-scale that follows immediately after, as shown in Fig.~\ref{fig:4}a. The output field displays a similar two-photon structure except that the second peak is relatively stronger in comparison to the case of the input field (see Fig.~\ref{fig:4}b).

Time-resolved intensity correlation maps $G^{(2)}(t_1,t_2)$, as shown in Figs.~\ref{fig:4}c,d provide insights into the processes occurring for this case of two incident photons. For the input field, one finds two-photon correlations along the horizontal and vertical axes of the map, which originate from temporally-orthogonal events---one photon first, and a second photon after. However, the output field also presents two-photon correlations along the diagonal, which are events of detecting two photons simultaneously. This correlated two-photon wave packet shows a measured decay time of $\tau'\,{=}\,108(6)$~ps, extracted from the exponential fit of the decay for $t_1\,{=}\,t_2$. This corresponds to approximately half the spontaneous emission decay time $\tau\,{=}\,227$~ps of the artificial atom (see Fig.~\ref{fig:1}b), a scaling factor that we also predict theoretically (see Fig.~\ref{fig:4}d). This significant observation explains the underlying mechanism: within all the complex quantum non-linear processes at play, sometimes one photon arrives at the artificial atom and excites it, and a subsequent photon arriving at the excited atom leads to the generation of two photons sharing the same temporal mode. These observations indicate stimulated emission processes~\cite{Rephaeli.2012} between only two quanta of light. This represents a fundamental form of mediated photon-photon interaction in a quantized field.

\subsection*{Discussion}
In this work, we reported the first observation of quantum light interacting with an atomic system in a solid-state platform. We started by employing a QD-cavity device that produces a high-fidelity single photon with high fiber-efficiency. Indeed, device efficiency has been key in allowing the observation of subtle changes in photon statistics within such processes, a task previously out of experimental reach. We then used a setup with fast and active path control that allows to send resonance fluorescence back to the quantum emitter and later observe and analyze the results of their interaction.

We studied different aspects of this process. First, we observed an artificial atom being resonantly excited by one photon. Here, we demonstrated the coherent nature of this process by observing single-photon interference mediated by an atom. An interesting aspect of this scenario is the observation that one photon is sufficient to excite the artificial atom, followed by the re-emission of a later photon. Moreover, we performed HOM experiments and showed that the output field produced at consecutive interaction instances remains indistinguishable. We also observed non-linear aspects of the light-matter interface acting on non-classical wave packets with one and two photons. We find that photon statistics change as a result of the interaction, with the output field containing more relative two-photon terms, which represents the first observation of photon-number dependent sorting occurring on non-classical input light.

Time-resolved intensity correlation maps also revealed a prime non-linear aspect of a quantum light-matter interface: stimulated emission. We observe that the output field contains simultaneous two-photon correlations, which are not present in the input non-classical field. We argue that these events occur when one resonant photon excites the atom, and a subsequent resonant photon stimulates the emission of a photon in the same spatio-temporal mode. The correlated output mode is a result of effective photon-photon interactions introduced in the highly non-linear medium.

We believe that our observations open a novel path for exploring light-matter interactions in a fully quantized system and will encourage a number of follow-up studies. For instance, one can envision performing this protocol in a sequential manner, where the output field resulting from the process is used again as input to observe a field resulting from multiple interactions, thus exploring the limits of the light-matter interface and answer questions such as up to which point the process can remain coherent or what types of non-classical light fields can be achieved asymptotically. Moreover, a QD emitting a temporal stream of photons entangled, e.g., in their polarization~\cite{entpairs:liu19} or photon-number~\cite{pnent:wein22}, could be used in a similar protocol, using either the same or a different QD, to entangle spatio-temporally multiple quantum emitters without requiring remote two-photon interference protocols or to create complex states of multi-partite entangled quantum light-matter systems.

\bibliography{biblio_general.bib}

\subsection*{Methods}
The InAs/GaAs QD investigated is coupled to a GaAs/AlAs distributive Bragg reflector that form a micropillar.
The device is connected to a surrounding diode structure, allowing for electrical tuning through the Stark effect. 
For the QD under study, the trion state is tuned into resonance with the cavity mode by applying $0.75$~V. 
The Purcell enhancement results in a detected radiation lifetime of $\tau\,{=}\,227~\text{ps}$ under $\pi$-pulse laser excitation, see Supplementary Information, with the QD transition emitting single photons at a wavelength of $922.1$~nm.
All the measurements are performed with the sample inside a closed-cycle cryostat (AttoDry800, Attocube) at a temperature of 4~K. To focus the excitation onto the sample and to collimate the output mode, a single aspheric lens with a numerical aperture of $0.68$ is installed inside the chamber. 

The initial excitation of the QD is performed using a pulsed Ti:Sapphire laser (Tsunami, Spectra Physics) with a pulse repetition rate of $80$~MHz. 
For pump preparation, we shape the laser pulse by spectral filtering with a 4$f$ system. 
The pulse length is set to ${\sim}25$~ps, assuming a transform-limited Gaussian pulse duration. 
A cross-polarized configuration is used for excitation and collection. In the collection path, linearly polarized laser light is suppressed using a set of wave plates and a Glan-Taylor polarizer with an extinction coefficient of more than seven orders of magnitude. The measured single-photon collection efficiency in the single-mode fiber is ${\sim}28.7~\%$~\cite{h3GHZ:Cao23}, including all losses from excitation to detection. 
Note that in the case of trion states, which have a circularly polarized transition, the resonant excitation scheme inherently introduces $50~\%$ loss since linearly polarized light is collected. 

For the preparation of non-classical excitation of a quantum emitter, the QD is excited resonantly by the pulsed laser, generating a temporal sequence of single photons that is then coupled to a single-mode fiber and sent to a broadband electro-optic modulator (EOM, QUBIG GmbH). 
At this stage, single photons, which serve as non-classical input, are deterministically sent back to the QD. 
To ensure that only one type of excitation is present at a given time, the delay of the returning single photons is adjusted accordingly such that the input and output signal are separated by a delay of ${\sim}3$~ns. 
The scattered Fock state then passes through the collection path and is actively routed by the EOM to a Hanbury Brown and Twiss (HBT) setup consisting of a balanced fiber beam splitter (BS) with two superconducting nanowire single-photon detectors (SNSPDs, Single Quantum) connected to a correlation unit (Time Tagger X, Swabian Instruments). 
The SNSPDs have a system detection efficiency of $86\%$, a timing jitter of $30$~ps, and a dark count rate of ${<}\,10$~Hz.
All devices are phase-locked to the pulsed laser excitation via a field programmable gate array (FPGA), which operates the EOM at a 50\% duty cycle at a frequency of $8$~MHz and serves as a reference clock for the measurement with a repetition rate of $8$~MHz. 

For the data analysis, the raw time tags are recorded which allows the histogram and correlation measurements to be carried out in post-processing. 
Thus, the raw time tags can be analyzed in a time-gated manner for each signal. 
For the time-resolved evaluation of the multi-photon probability $P_{m}(t_\text{start})$ shown in Fig.~\ref{fig:2}, the gating is defined by a start time $t_{\text{start}}$ and stop time $t_{\text{stop}}$ (in reference to the $8$~MHz clock signal produced by the FPGA), and all events occurring outside of the gating window $\Delta t_{\text{gate}}\,{=}\,t_{\text{stop}}\,{-}\,t_{\text{start}}$ are disregarded.
Correlating the gated channels then gives access to a time-gated $g^{(2)}(0, t_{\text{start}})$ value from which $P_{m}(t_1)$ is inferred~\cite{Vyvlecka.2023}. Increasing $t_{\text{start}}$ step-wise (while $t_{\text{stop}}$ is fixed) excludes the early fraction of the temporal profile and reveals a significant decrease in $P^{(m)}$ when the rising edge is omitted. Notably, the drop is much sharper for Fig.~\ref{fig:2}c (top) than for Fig.~\ref{fig:2}c (bottom).

Our theoretical simulations follow a cascaded master equation approach in which two identical subsystems with lifetime $\tau\,{=}\,\Gamma^{-1}\,{=}\,227$ ps are considered. The first system is driven by a Gaussian pulse with a FWHM of $\tau_p=25$~ps, leading to the output photon flux $\propto \Gamma \langle \sigma^{1}_+\sigma_-^{1}\rangle(t)$, shown in Fig.~1b. Here, $\sigma_\pm^{1}$ and $\sigma_\pm^{2}$ are the usual atomic ladder operators for the first and second system, respectively. The second system's output flux $\langle b^\dagger_2b_2\rangle(t)$, also shown in Fig.~1b, results from the output field operator $b_2(t)$, which is a weighted sum of the input field operator $\propto\sqrt{\Gamma}\sigma_-^{1}(t)$ and a system's response operator $\propto\sqrt{\Gamma}\sigma_-^{2}(t)$. To account for experimental imperfections, two different dephasing mechanisms were considered, and to account for the detector time resolution, a Gaussian detector jitter with a FWHM of $60$~ps was considered. 

\subsection*{Acknowledgments}
The authors thank Michael Trupke, Peter Rabl, Yuri Minoguchi and Juan Jos\'e Garc\'ia-Ripoll for helpful discussions, and Virginia Oddi and Simone Piacentini for help in initial steps of the project. This research was funded in whole, or in part, from the European Union’s Horizon 2020 and Horizon Europe research and innovation programme under grant agreement No 899368 (EPIQUS), No. 899354 (SuperQuLAN), and No 101135288 (EPIQUE), the Marie Sklodowska-Curie grant agreement No 956071 (AppQInfo), and the QuantERA II Programme under grant agreement No 101017733 (PhoMemtor); from the Austrian Science Fund (FWF) through [10.55776/F71           $]$ (BeyondC), [10.55776/FG5  $]$ (Research Group 5), and [10.55776/COE1  $]$ (Quantum Science Austria); by the Polish National Science Center (NCN) “Sonatina” project CARAMEL with number 2021/40/C/ST2/00155. The financial support by the Austrian Federal Ministry for Digital and Economic Affairs, the National Foundation for Research, Technology and Development and the Christian Doppler Research Association is gratefully acknowledged. 

\newpage
\clearpage

\renewcommand{\thefigure}{S\arabic{figure}}
\setcounter{section}{0}
\renewcommand{\thesection}{\Alph{section}}
\onecolumngrid
\setcounter{figure}{0}
\section*{Supplementary Information}
\subsection{Experimental setup}
\begin{figure}[htp!]
    \centering
    \includegraphics[width=0.8\textwidth]{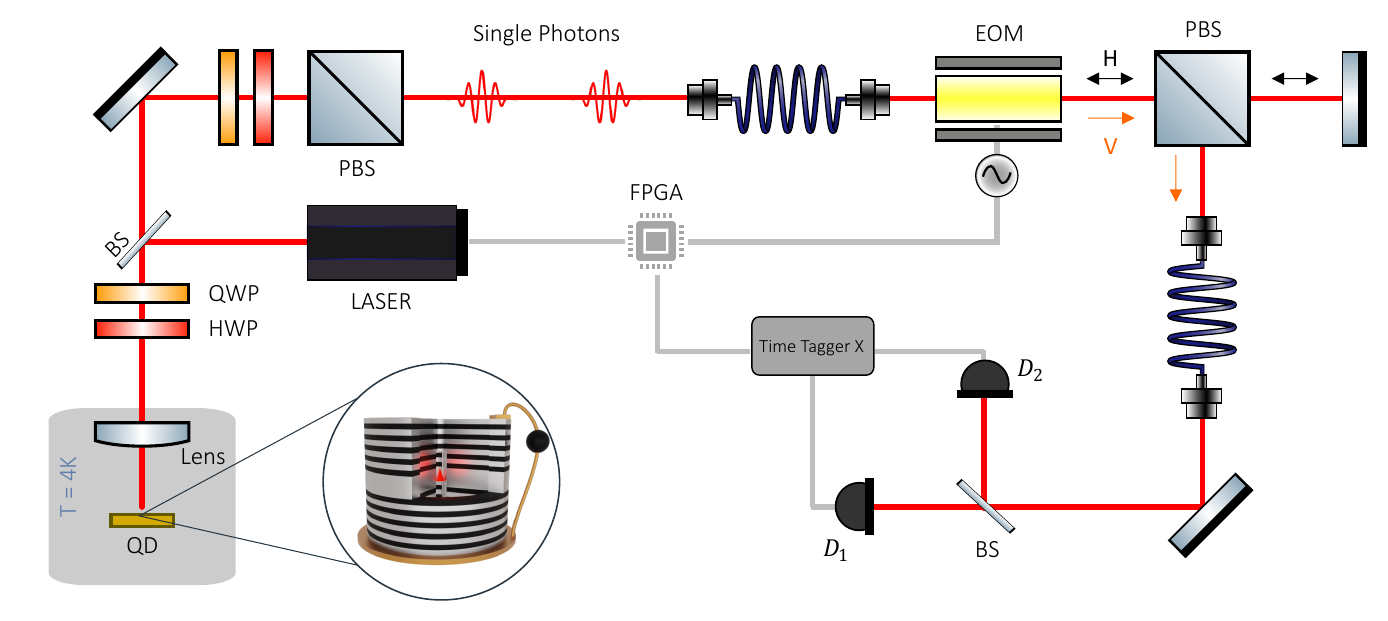}
    \caption{\textbf{Experimental setup.} A pulsed laser is used to excite the QD resonantly in a cross-polarized configuration. In the collection path, the laser is suppressed, and single photons are coupled into the collection fiber. In the following stage, an EOM is used to control the sequence in which single photons are sent back to the QD and light after the interaction with the QD is collected for analysis in a HBT setup.} 
\label{fig:setup_reex}
\end{figure}

The experimental setup for non-classical excitation of a QD is presented in Fig.\ref{fig:setup_reex}. 
The setup includes a cross-polarization excitation scheme, an active stage for sending single photons back to the emitter, and an analysis setup. 
The QD is first excited using resonant laser pulses. 
The sequence of single-photon time bins is then collected in a single-mode fiber, and the polarization state is set to horizontal polarization using fiber polarization control. 

During the first cycle, the EOM is in the off-state, which preserves the polarization state of the single photons such that light is transmitted through the PBS, is reflected by a mirror, and is coupled in reverse into the collection fiber. 
Thus, during the off-phase of the EOM, single-photon time bins return to the emitter. The scattered output signal after interaction with the QD is coupled to the collection fiber in the forward direction and sent to the EOM.

Following this, the EOM is set to the on-state, switching the polarization state of the light to vertical polarization. The light reflects at the PBS and is collected in the output fiber and sent to an HBT setup. 
This process allows simultaneous observation of the initial single-photon stream and the output signal, ensuring identical conditions for both excitation schemes. 
By recording the raw time-tag stream of each detector and the reference clock, both signals can be analyzed independently by time gating each signal in post processing. 
 
\subsection{Data set for different excitation parameters}  
\begin{figure}[htp!]
		\centering
        \includegraphics[width=1.0\textwidth]{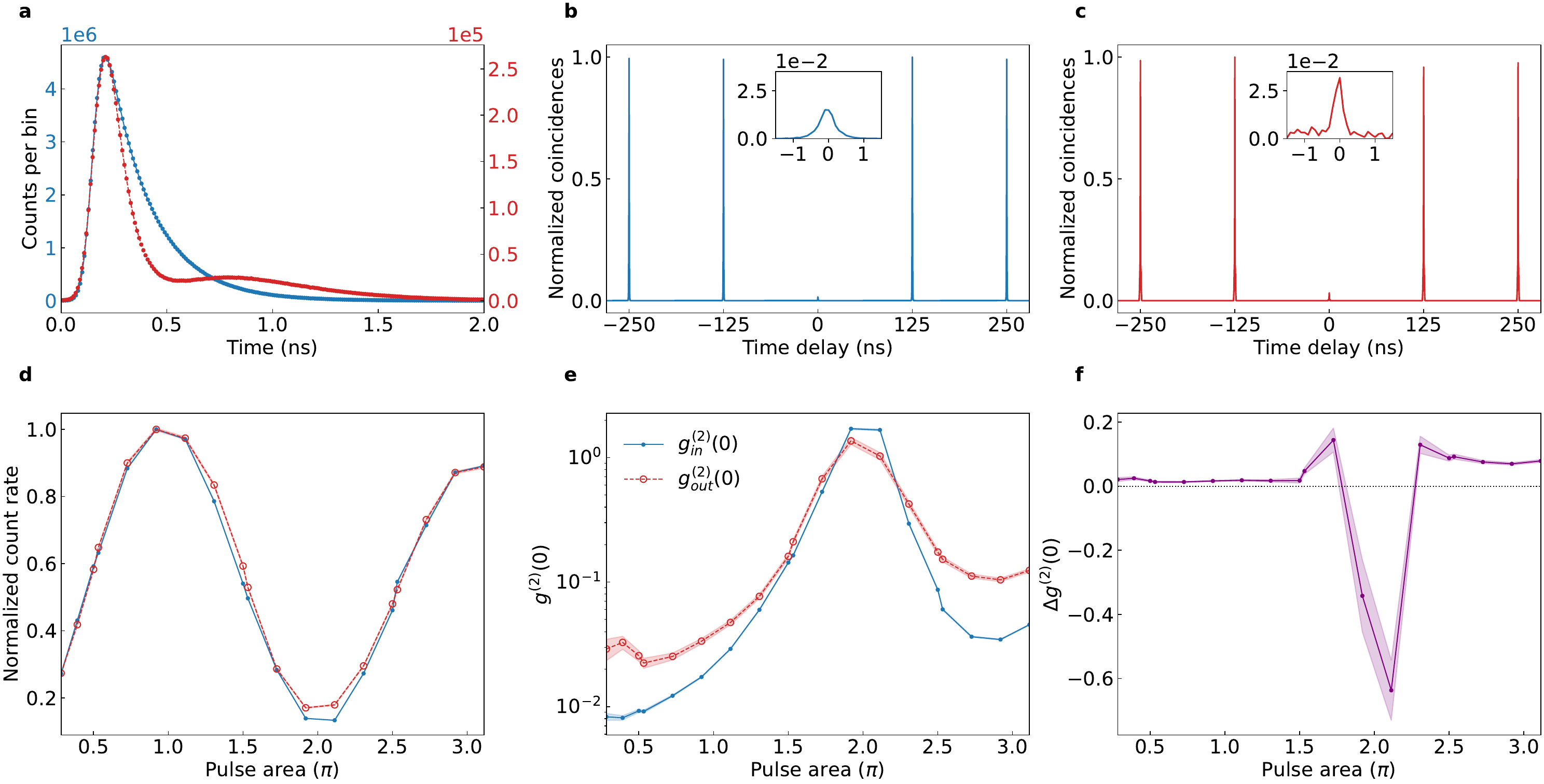}
		\caption{\textbf{Additional data set.} \textbf{a}, The time histograms shows single-photon emission and Fock state scattering. \textbf{b}, Normalized second-order auto-correlation function of the input field with a zero delay value of $g_{\text{in}}^{(2)}(0)=1.73(2)\%$ close to $\pi$-pulse excitation. \textbf{c}, than the normalized second-order auto-correlation function of the output field with a zero delay value of $g_{\text{out}}^{(2)}(0)=3.4(2)\%$. \textbf{d}, Measured Rabi oscillations for the input signal (blue) and the measured output count rate (red), that shows an increased transmitted photon flux for higher multi-photon contributions. \textbf{e}, Input photon statistic for increasing pulse area of the classical driving field and resulting photon statistic of the output field. \textbf{f}, Difference $\Delta g^{(2)}(0)\,{=}\,g^{(2)}_\text{out}(0){-}g^{(2)}_\text{in}(0)$ in photon statistics of input and output signal.}
	\label{fig:additional_may}
\end{figure}

For the measurements presented in the main text, we chose a laser pulse duration of ${\sim}25$~ps such a duration enhances the two-photon probability at the beginning of the lifetime because of reexcitation during the duration of the laser pulse \cite{Fischer:2017zr}. 
An additional set of measurements for a shorter laser pump pulse is given in Fig.~\ref{fig:additional_may}. 
For the additional data set, two-photon contributions are suppressed by setting the laser pulse length of ${\sim}18$~ps, that enhances single-photon purity while maintaining a high count rate.

The time histograms in Fig.~\ref{fig:additional_may}\textbf{a} indicate that the dynamics of th e output signal, as discussed in the main text, are reproducible under similar conditions. 
The interaction of the QD with the shorter laser pulse reduces the probability of two-photon emission during pulse duration as it is shown in the inset of the correlation measurement in Fig.\ref{fig:additional_may}\textbf{b}. 
Here, the central peak does not exhibit the characteristic emission profile of re-excitation during the duration of the laser pulse compared to the measurement presented in the main text. 
This results in a higher single-photon purity of the first generation with a second-order auto-correlation function value of $g_{\text{in}}^{(2)}(0)\,{=}\,1.73(2)\%$. 
The correlation measurement of the second generation is plotted in Fig.\ref{fig:additional_may}\textbf{c}. 
The measurement has a second-order auto-correlation function value of $g_{\text{out}}^{(2)}(0){=3}.4(2)\%$, indicating that the single-photon purity of the second generation is also improved as compared to the measurement presented in the main text, while remaining a lower single-photon purity relative to the first generation.

\begin{figure}[htp!]
		\centering
        \includegraphics[width=0.8\textwidth]{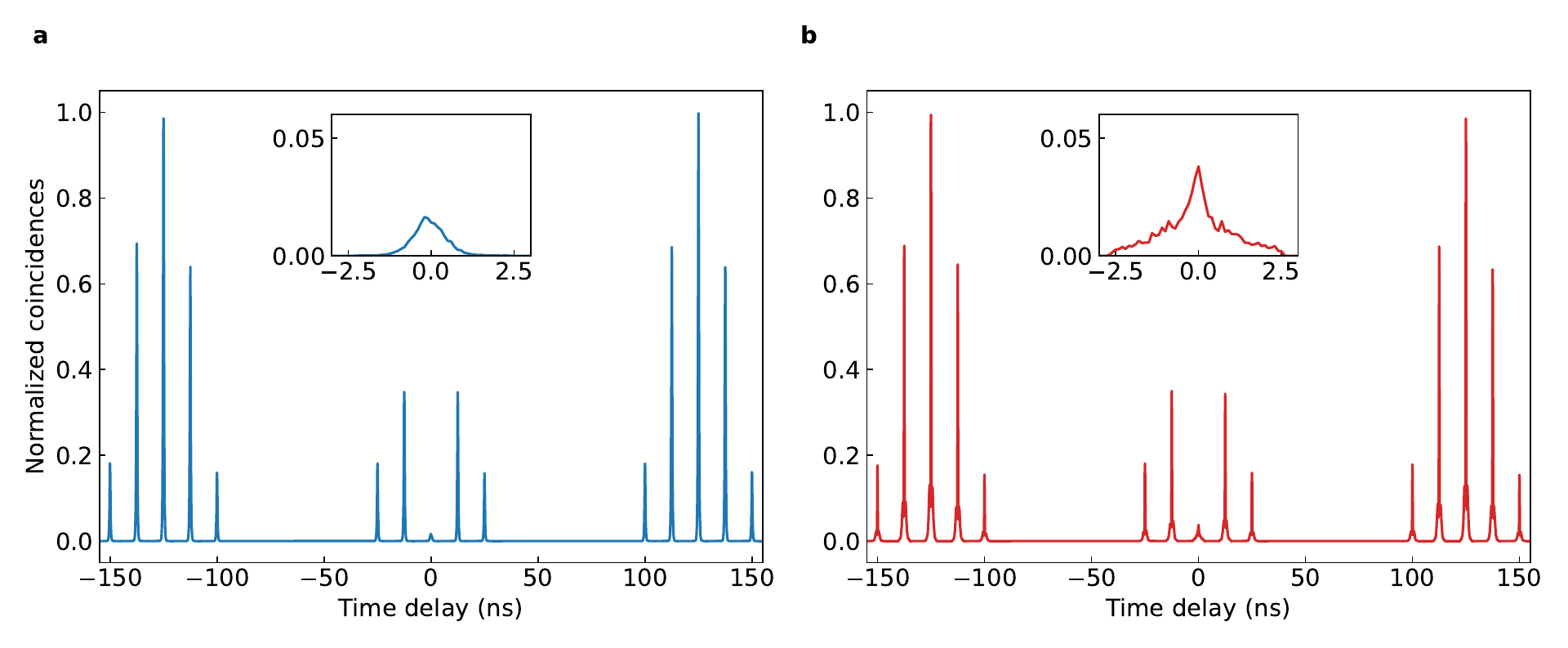}
		\caption{\textbf{HOM interference.} \textbf{a}, HOM measurement at $\pi$-pulse laser excitation. \textbf{b}, HOM measurement for non-classical excitation.} 
	\label{fig:09hom}
\end{figure}

Further, we investigate Hong–Ou–Mandel (HOM) interference of each signal. The measurement results are presented in Fig.~\ref{fig:09hom}. The raw HOM visibility of the initial single-photon signal generated by laser excitation is $V_{\text{in}}\,{=}\,90.0(2)\%$ (see Fig.~\ref{fig:09hom}\textbf{a}), yielding an indistinguishability value~\cite{HOM:Ollivier21} of $\mathcal{I}_\text{in}\,{=}\,93.3(2)\%$. The raw HOM visibility of the signal after interaction with the QD is $V_{\text{out}}\,{=}\,70.4(5)\%$ (see Fig.~\ref{fig:09hom}\textbf{b}), resulting in an indistinguishability value of $\mathcal{I}_\text{out}\,{=}\,76.4(6)\%$. The reported values do not account for unbalanced splitting ratios in the linear network.

The indistinguishability value is defined as $\mathcal{I}\,{=}\,\frac{V+g^{(2)}(0)}{1-g^{(2)}(0)}$~\cite{HOM:Ollivier21}. The raw HOM visibility is calculated as $V\,{=}\,1-A_0/A_1$, where $A_0$ is the integrated area of the central peak and $A_1$ is the mean integrated area of the peaks at a time delay of $12.5$~ns. The measurement uncertainties are estimated following Poissonian counting statistics. 

\subsection{Time histograms for classical and non-classical excitation}
\begin{figure}[htp!]
		\centering
        \includegraphics[width=0.8\textwidth]{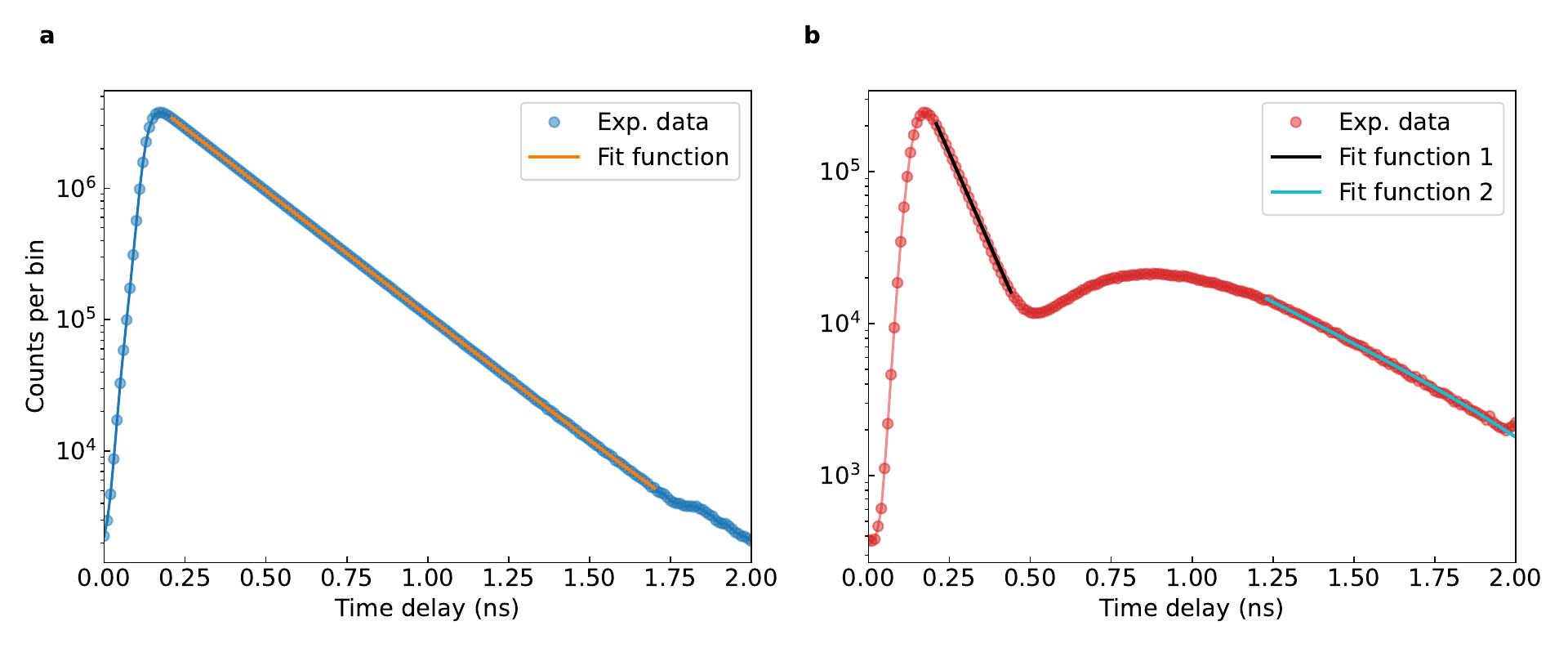}
		\caption{\textbf{Time histograms for single-photon emission and single-photon scattering.} \textbf{a}, The plot in a semilogarithmic scale shows the time-resolved fluorescence measurement of the QD state tuned into resonance with the cavity for $\pi$-pulse laser excitation. The system has a lifetime of $227$~ps, as evaluated by fitting a monoexponential decay to the experimental data. \textbf{b}, The plot shows the time-resolved histogram measurement of the scatter signal. The fits reveal a monoexponential decay of 90~ps for the first peak and $422$~ps for the second peak.} 
	\label{fig:pcov}
	\end{figure}

The time-resolved fluorescence measurements of the QD transition (trion) are displayed in Fig.~\ref{fig:pcov}\textbf{a}. The data represent the detected raw measurements with a resolution of $10$~ps (binwidth) and without corrections for a detector jitter of $30$~ps. Figure \ref{fig:pcov}\textbf{b} shows the detected time-resolved dynamics of the interaction between a single-photon wave packet and the emitter. Note that the dip and the attributed decay of the first peak originate from single-photon interference and absorption, while the second decay indicates the decay rate of the emission after the excitation of the emitter by a single photon~\cite{Npump:Combes12}.

\subsection{Theoretical model}
\subsubsection*{Cavity QED system coupled to a waveguide}
We model the system as a two-level system (TLS) interacting with two orthogonally polarized cavity modes (H and V) via the usual Jaynes-Cummings interaction, whose Hamiltonian reads ($\hbar=1$)~\cite{Auffeves2007, Antoniadis2022}
\begin{equation}
H_{\rm JC}= \frac{\omega_0}{2} \sigma_z+\omega_{\rm H}a^\dagger_{\rm H} a_{\rm H}+ \omega_{\rm V}a^\dagger_{\rm V} a_{\rm V} +i g_{\rm H}( \sigma_+ a_{\rm H} - \sigma_-a_{\rm H}^\dagger) + g_{\rm V}( \sigma_+ a_{\rm V} + \sigma_-a_{\rm V}^\dagger).
\end{equation}
Here, $\omega_0$ and $\sigma_{z,+,-}$ are the frequency and Pauli operators characterizing the TLS, $\omega_{\rm H,V}$ and $a^\dagger_{\rm H,V}$ ($a_{\rm H,V}$) are the frequencies and creation (annihilation) operators characterizing the cavity modes, and $g_{\rm H,V}$ are the light-matter coupling strengths. The one-sided cavity is coupled to the continuum with a total photon decay rate $\kappa$. In the ``bad cavity" limit, $\kappa \gg g_{\rm H,V}$, the cavity modes can be adiabatically eliminated, and the dynamics of TLS is then governed by the Lindblad master equation
\begin{equation}\label{Eq:UndrivenME}
\dot{\rho}=\mathcal{L}_0 \rho= -i\delta_L[\sigma_z,\rho]/2+ \Gamma D[\sigma_-]\rho + \gamma_d D[\sigma_z]\rho, 
\end{equation}
where $\rho(t)$ is the density matrix, $\delta_L=\omega_0-\omega_L$ is the detuning determined by the frequency of the rotating frame $\omega_L$, and $D[X]\rho=X\rho X^\dagger -\frac{1}{2}(X^\dagger X \rho + \rho X^\dagger X)$ is the Lindblad super-operator describing relaxation ($\sigma_-$) or dephasing ($\sigma_z$). On the one hand, the total decay rate of the TLS $\Gamma=\Gamma_{\rm H}+\Gamma_{\rm V}+\gamma^*$ accounts for losses via the cavity modes ($\Gamma_{\rm H,V}= 4g_{\rm H,V}^2/[\kappa (1+4\delta_{\rm H,V}^2/\kappa^2)]$ with $\delta_{\rm H,V}=\omega_0-\omega_{\rm H,V}$) or via other modes in the system ($\gamma^*\ll \Gamma$). On the other hand, $\gamma_d\ll \Gamma$ accounts for a phenomenological dephasing rate. 

For the sake of simplicity, in the following we consider $g_{\rm H}=g_{\rm V}$, $\delta_{\rm H}=\delta_{\rm V}=0$ and $\gamma^*=0$. As a result, $\Gamma_H=\Gamma_V=\Gamma/2$. Furthermore, we assume our cavity to be one sided. Then, the following input-output relations hold
\begin{equation}
b^{\rm H,V}_{out}=b_{in}^{\rm H,V}+\sqrt{\Gamma_{\rm H,V}}\,\sigma_-= b_{in}^{\rm H,V}+\sqrt{\Gamma/2}\,\sigma_-,
\end{equation}
where $b^{\rm H,V}_{in} $ and $b^{\rm H,V}_{out}$ are the input and output field operators, respectively. Note that in the absence of a horizontally-polarized input field, $\langle b_{in,\rm H}\rangle=0$, the measured output field flux $N(t)=\langle b_{out,\rm H}^\dagger(t)b_{out,\rm H}(t)\rangle$ is proportional to the probability that the TLS is in the excited state $P_{e}(t)=\frac{1}{2}(1+\langle\sigma_z(t)\rangle)$, which can be calculated via the density matrix $\rho(t)$ as $\langle X\rangle ={\rm Tr}(\rho (t)X) $. 

\subsubsection*{Laser driving and first-generation photon}
In our system, one of the cavity modes (V) is used to drive the TLS, while the other one (H) is used for measurement. We model the driving laser as a Gaussian pulse leading to an input field~\cite{DeSantis:2017aa,Fischer:2017zr}
\begin{equation}
\langle b^{\rm V}_{in}\rangle(t)=\sqrt{n_{L}}\bigg(\frac{4 \ln(2)}{\pi \tau_p^2}\bigg)^{1/4} \exp\Big(-\frac{2\ln(2) (t-t_c)^2}{\tau_p^2}\Big)
\end{equation}
where $\tau_p$ is the pulse width, $t_c$ is the instant characterizing its central point, and $n_L$ corresponds to the number of photons in the pulse, proportional to the laser intensity. This translates into a driving term in the effective TLS Hamiltonian, which in a frame rotating with the laser frequency $\omega_L$ reads
\begin{equation}
H(t)=\frac{\Omega (t)}{2}(\sigma_++\sigma_-),
\end{equation}
where $\Omega(t)=4g_{\rm V} \langle b^{\rm V}_{in,+}\rangle(t)/\sqrt{\kappa}$ is the Rabi frequency. It is convenient to express the Rabi frequency $\Omega(t)$ in terms of the pulse area $A$ as
\begin{equation}
\Omega(t)=A\bigg(\frac{2 \ln(2)}{\pi \tau_p^2}\bigg)^{1/2} \exp{(-2\ln(2) (t-t_c)^2/\tau_p^2)},
\end{equation}
ensuring $\int_0^\infty dt' \Omega(t')=A$ and removing dependance on parameters $\kappa$, $g_{\rm V}$ or $n_L$. Furthermore, we include a phonon-induced dephasing term as in Ref.~\cite{Fischer:2017zr}. Then, the Lindblad master equation for the first generation photon looks
\begin{equation}
\dot{\rho}=\mathcal{L}_\Omega \rho=-i[H(t),\rho] + \Gamma D[\sigma_-]\rho + \gamma_d D[\sigma_z]\rho + B \Omega^2(t) D[\sigma_z]\rho,
\end{equation}
while the measured output flux is
$N_1(t)\propto \langle b_{out,\rm H}^\dagger(t)b_{out,\rm H}(t)\rangle=\Gamma P_e(t) /2$.

\subsubsection*{Single-photon driving and second-generation photon}

In the experiment, the horizontally polarized light that comes out, $b_{out,1}^{\rm H}$, is reflected back to the cavity, and becomes the input field $b_{in,2}^{\rm H}$ at a later time, i.e. $b_{in,2}^{\rm H}=\sqrt{\eta_{\rm loss}}\,b_{out,1}^{\rm H}$. Here, $\eta_{\rm loss}$ accounts for the fraction of photons that survive after traveling through the optical setup back into the cavity. As the light entering the cavity is not coherent, we cannot represent it as a Hamiltonian term. Instead, we make use of a cascaded master equation approach~\cite{gardiner00}, where the output of a first system, driven by a laser pulse and with Pauli operators $\sigma_{\alpha}^{(1)}$ (with $\alpha{=}\{+,-,z\}$), is used as the input for a second system, with Pauli operators $\sigma_{\alpha}^{(2)}$. The full master equation then reads
\begin{equation}\label{Eq:CME}
\dot{\rho}=(\mathcal{L}^{(1)}_\Omega +\mathcal{L}^{(2)}_0  )\rho - \frac{\sqrt{\eta_{\rm loss}}}{2}\Gamma\big([\sigma_+^{(2)},\sigma_-^{(1)}\rho] +[\rho\sigma_+^{(1)},\sigma_-^{(2)}] \big).
\end{equation}
The horizontally polarized output field operator for the second system now reads
\begin{equation}
b^{\rm H}_{out,2}=b_{in,2}^{\rm H}+\sqrt{\Gamma/2}\, \sigma_-^{(2)}.
\end{equation}
As in reality our cavity is not perfectly one-sided, only a fraction of the light received by the atom will come out from the out mode $b^{\rm H}_{out,2}$. We model this by assuming a percentage $\eta_{\rm re}$ of the input light gets reflected right before entering the cavity. Then this fraction destructively interferes with the output light field $b^{\rm H}_{out,2}$. The modified measured field operator then reads
\begin{equation}
b^{\rm H}_{out,2}=\sqrt{\Gamma/2} [(\sqrt{\eta_{\rm loss}} -\sqrt{\eta_{\rm re}})\sigma_-^{(1)}+ \sigma_-^{(2)}]
\end{equation}
where $\eta_{\rm loss}=(1-\eta_{\rm re})\eta'_{\rm loss}$. 

\subsubsection*{Two-photon correlation functions}

The measured two-time correlation function $G^{(2)}(t_1,t_2)=\langle b^\dagger_{out}(t_1)b^\dagger_{out}(t_2) b_{out}(t_2)b_{out}(t_1)\rangle$ is equivalent to $G^{(2)}(t,t+\tau)$, where $\tau=|t_2-t_1|$, thus, we define $\overline{g^{(2)}(0)}$ as
\begin{equation}
    \overline{g^{(2)}(0)}=\frac{2\int_{0}^{\infty} dt\int_{0}^{\infty} d\tau G^{(2)}(t,t+\tau)}{|\int_{0}^{\infty} dt \langle b^\dagger_{out}(t)b_{out}(t)\rangle|^2},
\end{equation}
Under the previously considered approximations, the two-time correlation function for the first generation signal is
\begin{eqnarray}
G^{(2)}(t,t+\tau)= \frac{\Gamma^2}{4}  {\rm Tr}(\sigma_+^{(1)} \sigma_-^{(1)} \, \mathcal{U}^{(\tau)}[\sigma_-^{(1)} \mathcal{U}^{(t)}[\rho_0] \sigma_+^{(1)}]),
\end{eqnarray}
while, for the second generation, this is
\begin{eqnarray}
G^{(2)}(t,t+\tau)=\frac{\Gamma^2}{4} {\rm Tr}(S_+ S_- \, \mathcal{U}^{(\tau)}[S_-\mathcal{U}^{(t)}[\rho_0] S_+]),
\end{eqnarray}
where $S_\pm=(\sqrt{\eta_{\rm loss}} -\sqrt{\eta_{\rm re}})\sigma_\pm^{(1)}+ \sigma_\pm^{(2)}$. Here, $\rho_0$ is the initial state $\rho_0=|{\rm g}\rangle\langle {\rm g}|_1\otimes |{\rm g}\rangle\langle {\rm g}|_2$, and $\mathcal{U}^{(\tau)} $ is the time-evolution super-operator associated to the cascaded master evolution in Eq.~(\ref{Eq:CME}).  

\subsubsection*{Numerical simulation and relevant parameters}

The dynamics of our system is described by the cascaded master equation introduced in Eq.~(\ref{Eq:CME}). The free parameters are selected so as to maximise agreement with the experimental results. In particular, the value of the decay rate $\Gamma$ is chosen $\Gamma^{-1}=227$ ps, and the dephasing rates $\gamma_d=0.035 \Gamma$ and $B=1.3\times 10^{-4}\Gamma^{-1}$. Also, the reflection parameter $\eta_{\rm re}$ is chosen to be $\eta_{\rm re}=0.05$, while we find the studied observables independent of the loss parameter $\eta'_{\rm loss}$. Furthermore, the pulse parameters are chosen to be $\tau_p=25$ ps and $t_c=200$ ps. Finally, all time-dependent observables, e.g. $N(t)$, are assumed to have a finite time resolution limited by the detector's jitter, which is assume to be Gaussian with a full width at half maximum of $60$ ps.
\end{document}